\documentclass[11pt,a4paper,onecolumn,draft]{IEEEtran}

\usepackage{amsmath}
\usepackage{amssymb}
\usepackage{graphicx}
\usepackage{xcolor}
\usepackage{latexsym}
\usepackage{cite}
\usepackage{url}

\newtheorem{corollary}{Corollary}
\newtheorem{definition}{Definition}
\newtheorem{example}{Example}
\newtheorem{lemma}{Lemma}
\newtheorem{theorem}{Theorem}

\newcommand\Bc   {{\cal B}}
\newcommand\code {{\cal C}}
\newcommand\Dc  {{\cal D}}

\newcommand\Sc  {{\cal S}}
\newcommand\Tc  {{\cal T}}
\newcommand\Uc  {{\cal U}}
\newcommand\Xc   {{\cal X}}
\newcommand\Yc  {{\cal Y}}

\newcommand\calC  {{\cal C}}
\newcommand\xhat {{\hat x}}
\newcommand\Xhat {{\hat X}}

\newcommand\E    {{\mathbb E}}

\definecolor{dg}{rgb}{0, 0.5, 0}

\begin{document}

\title{On the Maximum Size of Block Codes Subject to a Distance Criterion}

\author{
\IEEEauthorblockN{
     Ling-Hua Chang\IEEEauthorrefmark{1},
     Po-Ning~Chen\IEEEauthorrefmark{1},
     Vincent Y.~F.~Tan\IEEEauthorrefmark{2}\IEEEauthorrefmark{3}, 
     Carol Wang\IEEEauthorrefmark{2},
     \\
     Yunghsiang S.~Han\IEEEauthorrefmark{4} 
\thanks{\IEEEauthorrefmark{1}Institute of Communications Engineering \& Department of Electrical and Computer Engineering, National Chiao Tung University, Taiwan, R.O.C.}
\thanks{\IEEEauthorrefmark{2}Department of Electrical and Computer Engineering, National University of Singapore, Singapore}
\thanks{\IEEEauthorrefmark{3}Department of Mathematics, National University of Singapore, Singapore}
\thanks{\IEEEauthorrefmark{4}School of Electrical Engineering \& Intelligentization,  Dongguan University~of Technology, China}
\thanks{iamjaung@gmail.com, cwang.ecc@gmail.com, poning@faculty.nctu.edu.tw, vtan@nus.edu.sg,  yunghsiangh@gmail.com}
}
}

\maketitle

\begin{abstract} 
We establish a general formula for the maximum size of finite length block
codes with minimum pairwise distance no less than $d$.
The achievability argument involves an iterative construction of a set of radius-$d$ balls, each centered at a codeword. We demonstrate that the number of such balls that cover the entire code alphabet cannot exceed this maximum size. 
Our approach can be applied to codes $i)$ with elements over arbitrary code alphabets, and $ii)$ under
a broad class of  distance measures, thereby ensuring the generality 
of our formula.
Our formula indicates that the maximum code
size can be fully characterized by the cumulative distribution function of the distance
measure evaluated at two independent and identically distributed random codewords.
When the two random codewords assume a uniform distribution over the entire code alphabet, our formula recovers
and obtains a natural generalization of 
the Gilbert-Varshamov (GV) lower bound.
We also establish a general formula for the zero-error capacity of any sequence of channels. 
Finally, we extend our study to the asymptotic setting, where we establish
first- and second-order bounds on the asymptotic code rate subject to a normalized minimum
  distance constraint. 
\end{abstract}

\section{Introduction}
\pagestyle{plain}

Given a code alphabet $\Xc$,  the determination of the maximal size $M_n^\ast(d)$ of a block code $\code \subseteq \Xc^n$ with pairwise minimum distance no less than $d$ and block length $n<\infty$ has been a long-standing problem in information and coding theory. 
One can use $M^\ast_n(d)$ to obtain an upper bound of the expurgated error exponent \cite{expurgated-error-exponent} and also to characterize the capacity of a graph \cite{lovasz}.
Some well-known bounds on $M^\ast_n(d)$ include the linear programming upper bound \cite{MRRW77} and Gilbert-Varshamov (GV) lower bound~\cite{MRRW77, Moon05,Lint92}. Other famous 
upper bounds include the Singleton, Plotkin, and Elias bounds \cite{MS83}.
However,   these bounds are not tight in general. Since finite-length bounds are usually difficult to obtain, researchers have
focused on asymptotic analyses in which blocklength $n$ tends to infinity. One then
considers the limit  of the code rate $(1/n)\log M_n^\ast(d)$ subject to 
a normalized distance constraint $d/n\geq\delta$. Many asymptotic bounds have been derived;
see, for example \cite{MRRW77,LT86,EZ87,Vladut87, Svanstrom97,Moon05,SX05,GZ08,BBGS14}  and the references therein. 

A natural question then beckons. Can one derive a ``meta-result" concerning the maximum code size  subject to a fixed minimum distance  $M^\ast_n(d)$ that recovers some of the above-mentioned bounds as special cases? 
In \cite{turan}, using a graph-theoretic framework, Motzkin and Straus derived such a result which implies  an exact formula for $M^\ast_n(d)$ under the condition that   $\Xc$ is finite. See also a related result by Korn~\cite{korn}. It is then natural to ask if there exists an analogous result for more general code alphabets, e.g., uncountable alphabets.  This is precisely the purpose of this paper. Our contributions are as follows:

\begin{enumerate}
\item We propose an iterative construction of a set of balls, each centered at a codeword and of a fixed radius $d$. We then show that the number of such balls that cover the entire code alphabet cannot exceed the maximum code size subject to a fixed minimum distance $d$.
Consequently, we prove that $M_n^\ast(d)$ for an 
 {\em arbitrary} code alphabet 
can be completely
determined by the minimum probability (over all distributions) that two i.i.d.~random vectors $\Xhat^n$ and $X^n$ are at distance less than $d$ from each other, i.e., 
\begin{equation}
M_n^\ast(d)=\frac 1{\inf_{P_{X^n}}\Pr\big[\min\{\mu(\Xhat^n,X^n),\mu(X^n,\Xhat^n)\}<d\big]}, \label{eqn:intro_eq}
\end{equation}
where $\mu(\cdot,\cdot)$ is the (possibly asymmetric) distance measure.
Our formula can be used to recover and obtain 
a natural generalization of the GV bound. 
   
\item Based on our result, a general formula for the zero-error capacity~\cite{zero_error} of any sequence of channels can be obtained. 
\item Finally, we derive a general expression for the maximum asymptotic code rate $\frac 1n\log M_n^\ast(n\delta)$ as $n$ approaches infinity 
 such that the relative minimum distance is at least $\delta$. Our proof also characterizes
a second-order bound (cf.\ Lemma \ref{lemma2}) on the maximum asymptotic code rate.
\end{enumerate}

The term $\Pr[\min\{\mu(\Xhat^n,X^n),\mu(X^n,\Xhat^n)\} < d]$ can be regarded as the information spectrum \cite{shannon,Han03} (or cumulative distribution function) of the distance measure
$\min\{\mu(\Xhat^n,X^n),\mu(X^n,\Xhat^n)\}$. Posing the problem of finding $ M^\ast_n(d)$ as an optimization
problem over a continuous  variable (the distribution $P_{X^n}$) can potential yield  new approaches for finding   codes with improved performances over existing codes.

The rest of the paper is organized as follows.
The exact   formula for $M_n^\ast(d)$ is presented in Section~\ref{sec:mainresult}. This formula is then used to obtain a general formula for the zero-error capacity of a sequence of channels. 
A family of lower bounds to $M^\ast_n(d)$ is presented in Section~\ref{sec:implications}; also included here is the demonstration that the  finite length GV lower bound  can be recovered from our formula. 
Extensions to the asymptotic regime are studied in Section~\ref{sec5:asym}. 
Finally, open problems are discussed in Section~\ref{sec:conclusion}.

\section{Maximal Code Size Attainable under a Minimum Pairwise Distance}
\label{sec:mainresult}

We first introduce the notation used in this paper. 
An $(n,M)$-code over alphabet $\Xc^n$ denotes a set of $M$ codewords, 
each of which belongs to  $\Xc^n$ \cite{LC04}. 
A distance measure $\mu(\cdot,\cdot)$ is a real-valued function with domain $\Xc^n\times\Xc^n$ which satisfies
\begin{equation}
\label{dist-cond}
\mu( u^n,v^n)= \mu_{\min}\triangleq\min_{\hat{x}^n,x^n\in\Xc^n} \mu( \hat{x}^n,x^n )\quad \text{if} \quad u^n=v^n.
\end{equation}
Here, we do not require $\mu(\cdot,\cdot)$ to be symmetric or satisfy the triangle inequality but can be arbitrary as long as it admits its minimum from a point to itself.

An $(n,M,d)$-code $\code$ denotes 
an $(n,M)$-code with the minimum pairwise distance among codewords at least $d$, i.e.,
\begin{equation}
\min_{\xhat^n,\,x^n\in\code\text{ and }\xhat^n\neq x^n}\mu(\xhat^n,x^n)\geq d. \label{eq:min}
\end{equation}
The maximal code size $M_n^\ast(d)$ subject to a pairwise minimum distance $d$ is
  given by
\begin{equation}
M_n^\ast(d) \triangleq \max\left\{ M\in\mathbb{N}: \exists\, (n,M,d)\mbox{-code} \right\},\label{eqn:m_star}
\end{equation}
where $\mathbb{N}$ is the set of positive integers. 
For convenience, a code that satisfies \eqref{eq:min}
is referred to as a \emph{distance-$d$ code}. 
Throughout this paper, $\Xhat^n$ and $X^n$ denote two independent random variables with a common distribution $P_{X^n}$ over $\Xc^n$. 

\subsection{Distance Spectrum Formula of $M^\ast_n(d)$}

We now present a general  formula for the maximum size $M^\ast_n(d)$ of distance-$d$ codes
over an arbitrary code alphabet $\Xc$ (not necessarily  countable) and 
general distance measure $\mu(\cdot,\cdot)$. 


\begin{theorem}\label{thm-m-size} 
Fix an arbitrary code alphabet $\Xc$ 
and a distance measure $\mu(\cdot,\cdot)$ that 
satisfies \eqref{dist-cond}. For all $n\geq 1$ and $d>\mu_{\min}$,
we have 
\begin{IEEEeqnarray}{rCl}
M_n^\ast(d)=\frac 1{\inf_{P_{X^n}}\Pr\big[\min\{\mu(\Xhat^n,X^n),\mu(X^n,\Xhat^n)\}<d\big]}.
\label{eq:optimalcodesize}
\end{IEEEeqnarray}
\end{theorem}
\begin{IEEEproof} 
We first prove the validity of \eqref{eq:optimalcodesize} under the assumption that $M_n^\ast(d)<\infty$.
Its extension to $M_n^\ast(d)=\infty$ will be done next.

Subject to  the condition that $M_n^\ast(d)$ is finite, the equality in \eqref{eq:optimalcodesize} can be proved in two steps.
We first show that for every distribution $P_{X^n}$ over $\Xc^n$, the following inequality holds:
\begin{equation}
\Pr\big[\tilde\mu(\Xhat^n,X^n)<d\big]\geq \frac 1{M_n^\ast(d)},
\label{eq:basic}
\end{equation}
where for convenience, we denote $\tilde\mu(\xhat^n,x^n)\triangleq\min\{\mu(\xhat^n,x^n),\mu(x^n,\xhat^n)\}$; hence,
\begin{equation}
\inf_{P_{X^n}}\Pr\big[\tilde\mu(\Xhat^n,X^n)<d\big]\geq \frac 1{M_n^\ast(d)}.
\label{eq:basic-a}
\end{equation}
The proof is then completed by exhibiting a distribution $P_{X^{n\ast}}$ that results in equality in \eqref{eq:basic-a}; consequently, given that $M_n^\ast(d)$ is finite, the infimum 
 in~\eqref{eq:optimalcodesize} 
can be replaced by a minimum.
\begin{enumerate}
\item Achievability (\emph{Validation of \eqref{eq:basic} under finite $M_n^\ast(d)$}):  
Fix a distribution $P_{X^n}$ over $\Xc^n$ and an arbitrarily small $\epsilon>0$. Let 
\begin{equation}
a_1\triangleq\inf_{x^n\in\Xc^n}\Pr[X^n\in\Bc(x^n)],
\end{equation}
where $\Bc(x^n)\triangleq\{\xhat^n\in\Xc^n:\tilde\mu(\xhat^n,x^n)<d\}$.
Find an element $u_1^n$ in $\Xc^n$ such that 
$p_1\triangleq\Pr[X^n\in\Bc(u_1^n)]<a_1+\epsilon$. Note that the existence of $u_1^n$ is guaranteed by the definition of the infimum. Let 
\begin{equation}
a_2\triangleq\inf_{x^n\in\Xc^n\setminus\Bc(u_1^n)}\Pr[X^n\in\Bc(x^n)\setminus\Bc(u_1^n)].
\end{equation}
Find an element $u_2^n$ in $\Xc^n\setminus\Bc(u_1^n)$ such that 
$p_2\triangleq\Pr[X^n\in\Bc(u_2^n)\setminus\Bc(u_1^n)]<a_2+\epsilon$.
We repeat this procedure to obtain  
\begin{equation}
a_i\triangleq\inf_{x^n\in\Xc^n\setminus\cup_{j=1}^{i-1}\Bc(u_j^n)}\Pr[X^n\in\Bc(x^n)\setminus\cup_{j=1}^{i-1}\Bc(u_j^n)] \label{eqn:ai}
\end{equation}
and an $u_i^n$ in $\Xc^n\setminus\cup_{j=1}^{i-1}\Bc(u_j^n)$ 
with $p_i\triangleq\Pr[X^n\in\Bc(u_i^n)\setminus\cup_{j=1}^{i-1}\Bc(u_j^n)]<a_i+\epsilon$ for $i=3,4,\ldots,k$ until $\cup_{j=1}^{k}\Bc(u_j^n)$ covers the entire $\Xc^n$, i.e.,
$\Xc^n\setminus \cup_{j=1}^{k}\Bc(u_j^n)=\emptyset$
but $\Xc^n\setminus \cup_{j=1}^{k-1}\Bc(u_j^n)\not=\emptyset$.
Two observations are made:
$i)$  
$\{u_1^n,u_2^n,\ldots,u_k^n\}$ is a distance-$d$ code and hence by the definition of $M_n^\ast(d)$ and its assumed finiteness, $k\leq M_n^\ast(d)$ is a finite integer so the above procedure is repeated at most $M_n^\ast(d)$ times; $ii)$ $\sum_{j=1}^k p_j=1$.
Denoting $\Dc_i\triangleq\Bc(u_i^n)\setminus\cup_{j=1}^{i-1}\Bc(u_j^n)$
and noting $\Dc_i\cap\Dc_j=\emptyset$ for $i\neq j$ and $\Xc^n=\cup_{j=1}^k \Dc_j$ (i.e., $\{ {\cal D}_j \}_{j=1}^k$ is  a partition of $\Xc^n$),  
we can derive the following chain of inequalities:
\begin{IEEEeqnarray}{rCl}
\Pr[\tilde\mu(\Xhat^n,X^n)<d]
&=&\int_{\Xc^n}\int_{\Xc^n}\,{\bf 1}\{\tilde\mu(\xhat^n,x^n)<d\}\,\mathrm{d}P_{X^n}(\xhat^n)\,\mathrm{d}P_{X^n}(x^n)\\
&=&\sum_{j=1}^k\int_{\Dc_j}\int_{\Xc^n}\,{\bf 1}\{\tilde\mu(\xhat^n,x^n)<d\}\,\mathrm{d}P_{X^n}(\xhat^n)\,\mathrm{d}P_{X^n}(x^n)\\
&=&\sum_{j=1}^k\int_{\Dc_j}\int_{\Bc(x^n)}\,\mathrm{d}P_{X^n}(\xhat^n)\,\mathrm{d}P_{X^n}(x^n)\\
&\geq&\sum_{j=1}^k\int_{\Dc_j}\,a_j\,\mathrm{d}P_{X^n}(x^n)\label{eq:aj}\\
&=&\sum_{j=1}^k a_jp_j\label{eq:pj}\\
&>&\sum_{j=1}^k (p_j-\epsilon)p_j\label{eq:apj}\\
&=&\bigg(\sum_{j=1}^k p_j^2\bigg)-\epsilon\label{eq:sum}\\
&\geq&\frac 1k-\epsilon\label{eq:CSnew}\\
&\geq&\frac 1{M_n^\ast(d)}-\epsilon,\label{eq:last}
\end{IEEEeqnarray}
where 
${\bf 1}\{\cdot\}$ is the set indicator function;
\eqref{eq:aj} holds because
\begin{IEEEeqnarray}{rCl}
\inf_{x^n\in\Dc_j}\int_{\Bc(x^n)}\,\mathrm{d}P_{X^n}(\xhat^n)
&=&\inf_{x^n\in\Bc(u_j^n)\setminus\cup_{\ell=1}^{j-1}\Bc(u_\ell^n)}
\Pr[X^n\in\Bc(x^n)]\\
&\ge &\inf_{x^n\in\Bc(u_j^n)\setminus\cup_{\ell=1}^{j-1}\Bc(u_\ell^n)}
\Pr[X^n\in\Bc(x^n)\setminus\cup_{\ell=1}^{j-1}\Bc(u_\ell^n)]\\
&\geq&\inf_{x^n\in\Xc^n\setminus\cup_{\ell=1}^{j-1}\Bc(u_\ell^n)}
\Pr\big[X^n\in\Bc(x^n)\setminus\cup_{\ell=1}^{j-1}\Bc(u_\ell^n)\big]=a_j;
\end{IEEEeqnarray}
\eqref{eq:pj} follows from the definition of $p_j$; \eqref{eq:apj} holds since $p_j<a_j+\epsilon$;
\eqref{eq:sum} applies since $\sum_{j=1}^kp_j=1$; \eqref{eq:CSnew} is a consequence of the Cauchy-Schwarz inequality;\footnote{The Cauchy-Schwarz inequality
 can be used to assert that
$
1=\big(\sum_{j=1}^k 1\cdot p_j\big)^2\leq
\big(\sum_{j=1}^k 1^2\big)\big(\sum_{j=1}^k p_j^2\big)=k\sum_{j=1}^k p_j^2.
$} and the last inequality  in~\eqref{eq:last} follows from $k\leq M_n^\ast(d)$.
The proof of \eqref{eq:basic} is completed by noting that 
the above derivations hold for arbitrarily small $\epsilon$.

\item Converse (\emph{Equality of \eqref{eq:basic-a} under finite $M_n^\ast(d)$}):
Let $P_{X^{n\ast}}$ be the uniform distribution over a distance-$d$ code $\code^\ast$ that achieves $M_n^\ast(d)$. 
We then have
\begin{IEEEeqnarray}{rCl}
\Pr[\tilde\mu(\Xhat^{n\ast},X^{n\ast})<d] = \sum_{x^n\in\code^\ast}\big[P_{X^{n\ast}} (x^n) \big]^2 = \sum_{x^n\in\code^\ast}\frac 1{|\code^\ast|^2} =\frac 1{M_n^\ast(d)},
\label{eq:optimalcode}
\end{IEEEeqnarray}
where $|\code^\ast|$ denotes the cardinality of $\code^\ast$.
\end{enumerate}
The above two steps complete the proof of 
\begin{equation}
\min_{P_{X^n}}\Pr\big[\tilde\mu(\Xhat^n,X^n)<d\big]=\frac 1{M_n^\ast(d)}.
\end{equation}
subject to finite $M_n^\ast(d)$.

When $M_n^\ast(d)=\infty$, again, let $\code^\ast$ denote an infinite distance-$d$ code that achieves $M_n^\ast(d)$.
Then, any finite subset $\Sc$ of $\code^\ast$ is a distance-$d$ code.
Using a derivation similar to that leading to \eqref{eq:optimalcode}
gives that
\begin{IEEEeqnarray}{rCl}
\Pr[\tilde\mu(\Xhat^{n\circ},X^{n\circ})<d] = \frac 1{|\Sc|},
\end{IEEEeqnarray}
where $P_{X^{n\circ}}$ is the uniform distribution over $\Sc$.
As $|\Sc|$ can be made arbitrarily large,
\begin{equation}
\inf_{P_{X^n}}\Pr\big[\tilde\mu(\Xhat^n,X^n)<d\big]=0.
\end{equation}
This completes the proof.
\end{IEEEproof}

Some remarks concerning Theorem~\ref{thm-m-size} are in order.
First, the theorem can be applied to an arbitrary code alphabet and 
any distance measure satisfying \eqref{dist-cond}.
Its generality thus extends the study of the maximal code size of distance-$d$ codes
from the conventional finite code alphabets and the Hamming distance 
to, for example, $\Xc=[0,1)$ and the Euclidean distance (cf.~Example~\ref{example1}).

Secondly, the crux of the proof of Theorem~\ref{thm-m-size} 
is the observation that the entire space $\Xc^n$ can be covered by $k$ ``open" balls of radius $d$
with $k\leq M_n^\ast(d)$,\footnote{Here ``open" means a strict inequality is used to define the ball. }
 where the radius is defined via the distance $\tilde\mu(\cdot,\cdot)$. 
 In addition, the selection of the center $u_i^n$ of the next ball $\Bc(u_i^n)$ is chosen such that 
$p_i=\Pr[X^n\in\Bc(u_i^n)\setminus\cup_{j=1}^{i-1}\Bc(u_j^n)]$ is
$\epsilon$-close to its minimum possible value
and therefore $k$ can be made as large
as possible, ideally as close to $M_n^\ast(d)$ as possible. 
 
Thirdly, as noted by Korn~\cite{korn}, when the code alphabet 
$\Xc^n$ is finite, the optimization  problem $\inf_{P_{X^n}}\Pr[\tilde\mu(\Xhat^{n},X^{n})<d]$
corresponds exactly to the minimization of the quadratic form ${\bf p}\mathbb{A}  {\bf p}^{\tt T}$,
where ${\bf p}$ is the row vector formed by listing the probability masses of $P_{X^n}$ and
$\mathbb{A}$ is the corresponding $|\Xc|^n\times |\Xc|^n$ matrix with entries given by ${\bf 1}\{\tilde\mu(\xhat^n,x^n)<d\}$.
This quadratic optimization problem was considered by Korn~\cite{korn} in his study of 
the maximization of Gallager's lower bound for the zero-error capacity of discrete memoryless channels (DMCs)~\cite{Gallager}. 
The same solution can also be found in Motzkin and Strass' work~\cite{turan},
where the order of the maximal complete graph contained in a {\em finite graph} is considered. 
Here, instead of iteratively removing one codeword from any two codewords within distance $d$ until the size of the set of candidate codewords is reduced to  $M_n^\ast(d)$ as suggested by Korn's technique in \cite{korn},
we define a ``proper" notion of progress to iteratively add
codewords to a distance-$d$ code.
Specifically, we select a representative vector $u_i^n$ in some ``$\epsilon$-neighborhood'' defined as $\{u^n\in\Xc^n:\Pr[X^n\in\Bc(u^n)\setminus\cup_{j=1}^{i-1}\Bc(u_j^n)]<\inf_{x^n\in\Xc^n}\Pr[X^n\in\Bc(x^n)\setminus\cup_{j=1}^{i-1}\Bc(u_j^n)]+\epsilon\}$ for a given distribution $P_{X^n}$. 
This selection is repeated until the entire code alphabet can be covered by the union of radius-$d$ balls centered at $u_i$'s. 
The assumed finiteness of $M_n^\ast(d)$ ensures that the iterative selection will terminate.
Note that the proof of Theorem~\ref{thm-m-size} is not restricted to code alphabets that are finite (cf.~\cite{turan, korn}). In addition to being applicable to general arbitrary code alphabets, it provides a different perspective of the general formula in~\eqref{eq:optimalcodesize}.

\subsection{Relation of Theorem  \ref{thm-m-size} to the Zero-Error Capacity}

We now show that  Theorem \ref{thm-m-size} can be used to establish a general formula for the zero-error capacity~\cite{zero_error}  for arbitrary channels. This result complements  the  general formula for the  (vanishing error) capacity of arbitrary channels considered by Verd\'u and Han in~\cite{verdu-han}. 


\begin{definition}[Zero-error capacity]\label{definition2}
Let $\Omega_n$ be the maximum code size that can be transmitted error-free (i.e., with exactly zero error probability) over the channel $P_{Y^n|X^n}$. Then, the zero-error capacity for a sequence of channels {$\{P_{Y^n|X^n}\}_{n=1}^\infty$} is defined as
\begin{equation}
C_0\triangleq {\sup_{n\geq 1}}\;\frac 1n\log\Omega_n.
\label{eq:zero-error-capacity}
\end{equation}
\end{definition}


\begin{theorem}[General zero-error capacity]\label{theorem2}
The zero-error capacity for an arbitrary sequence of  channels {$\{P_{Y^n|X^n}\}_{n=1}^\infty$}  (not necessarily with countable alphabets) can be expressed as
\begin{IEEEeqnarray}{rCl}
C_0
&=&{\sup_{n\geq 1}}\;-\frac 1n\log \inf_{P_{X^n}}\Pr[\mu(\Xhat^n,X^n)=0],
\end{IEEEeqnarray}
where
\begin{IEEEeqnarray}{rCl}
\mu(\hat{x}^n,x^n)
&\triangleq&\begin{cases}
1,&(\exists~\Tc\subset\Yc^n)\ E(\xhat^n;\Tc)=E(x^n;\Yc^n\setminus\Tc)=1;\\
0,&\text{otherwise},
\end{cases}\label{d-zero}
\end{IEEEeqnarray}
and $E(x^n;\Tc)\triangleq \Pr[Y^n\in\Tc|X^n=x^n]$ for every measurable $\Tc$. 
\end{theorem}

Note that when $\Yc$ is finite, \eqref{d-zero} implies that 
$\mu(\xhat^n,x^n)=0$ if and only if there exists an $y^n$ such that
the channel $P_{Y^n|X^n}$ maps both $\xhat^n$ and $x^n$ to 
$y^n$ with positive probabilities.  
\begin{IEEEproof}
For $\xhat^n$ and $x^n$ with $\mu(\xhat^n,x^n)=1$,
we denote by $\Dc_{\xhat^n}(x^n)$ and $\Dc_{x^n}(\xhat^n)$ 
two subsets of $\Yc^n$ satisfying  
$E(\xhat^n;\Dc_{\xhat^n}(x^n))=E(x^n;\Dc_{x^n}(\xhat^n))=1$
and $\Dc_{\xhat^n}(x^n)\cap \Dc_{x^n}(\xhat^n)=\emptyset$.
Note that the existence of disjoint $\Dc_{\xhat^n}(x^n)$
and $\Dc_{x^n}(\xhat^n)$ is guaranteed by 
$\Dc_{\xhat^n}(x^n)\subset\Tc$ and
$\Dc_{x^n}(\xhat^n)\subset\Yc^n\setminus\Tc$,
where $\Tc$ is the set to make $\mu(\xhat^n,x^n)=1$ in 
\eqref{d-zero}.

Then, the next two steps verify that $\Omega_n$ in Definition~\ref{definition2} is equal to $M_n^\ast(1)$.
\begin{enumerate}
\item Achievability: To show that 
a code $\calC_n$ that achieves $M_n^\ast(1)$ can be decoded error-free,
we define
\begin{IEEEeqnarray}{rCl}
\Uc(\xhat^n)\triangleq \bigcap_{x^n\in\calC_n\,:\, x^n\neq\xhat^n}\Dc_{\xhat^n}(x^n)
\end{IEEEeqnarray}
and note that for every distinct pair of $\xhat^n$ and $x^n$ in the code, we have 
$\Uc(\xhat^n)\cap\Uc(x^n)=\emptyset$
because 
$\Dc_{\xhat^n}(x^n)\cap \Dc_{x^n}(\xhat^n)=\emptyset$.
As a result, the decoder that decodes $y^n$ to $x^n\in\calC_n$ if $y^n\in\Uc(x^n)$, and to an arbitrary codeword if $y^n\not\in\cup_{x^n\in\calC_n}\Uc(x^n)$, 
has zero error probability. 
Hence, $\Omega_n\geq M_n^\ast(1)$.

\item Converse: Suppose $\Omega_n> M_n^\ast(1)$. Let the code that achieves $\Omega_n$ be denoted by $\calC_n$. Then, there exist a partition of  disjoint subsets
$\{\Uc(x^n)\}_{x^n\in\calC_n}$ on $\Yc^n$ such that the decoder that decodes
$y^n$ to $x^n$ when $y^n\in\Uc(x^n)$ has zero error probability.
Therefore, $E(\xhat^n;\Uc(x^n))=0$ whenever $\xhat^n$ and $x^n$ are two distinct elements in $\calC_n$. This implies that $E(\xhat^n;\Uc(\xhat^n))=1$ since 
$\sum_{x^n\in\calC_n}E(\xhat^n;\Uc(x^n))=1$.
As a result of $E(\xhat^n;\Uc(\xhat^n))=E(x^n;\Yc^n\setminus\Uc(\xhat^n))=1$, 
we have $\mu(\hat{x}^n,x^n)=1$ for all distinct pairs of $\xhat^n$ and $x^n$ in $\calC_n$, 
implying $M_n^\ast(1)\geq\Omega_n$. Thus, a contradiction to the assumption that $M_n^\ast(1)<\Omega_n$ is obtained.
\end{enumerate}
By noting that $\mu(\xhat^n,x^n)=\mu(x^n,\xhat^n)$ and $\Omega_n =  M_n^\ast(1)$, the proof of 
the theorem is completed by noting that
\begin{IEEEeqnarray}{rCl}
C_0 &=&
{\sup_{n\geq 1}}\;\frac 1n\log M_n^\ast(1)\\
&=&{\sup_{n\geq 1}}\; -\frac 1n\log \inf_{P_{X^n}}\Pr[\mu(\Xhat^n,X^n)<1]\\
&=&{\sup_{n\geq 1}}\;-\frac 1n\log \inf_{P_{X^n}}\Pr[\mu(\Xhat^n,X^n)=0].
\end{IEEEeqnarray}
\end{IEEEproof}

The next corollary presents an alternative derivation of Korn's lower bound to the zero-error capacity for DMCs~\cite{korn}.


\begin{corollary}  \label{cor:dmc}
For DMCs with finite channel input alphabet $\Xc$ and finite channel output alphabet $\Yc$, 
\begin{equation}
C_0\geq-\frac 1n\log \inf_{P_{X}}\Pr\left[
\sum_{y\in \Yc}P_{Y|X}(y|\hat{X})P_{Y|X}(y|X)>0\right].
\end{equation}
\end{corollary}
\begin{IEEEproof}
We first note that for DMCs with finite $\Xc$ and $\Yc$, 
\eqref{d-zero} can be equivalently written as
\begin{IEEEeqnarray}{rCl}
\mu(\hat{x}^n,x^n)
&=&\begin{cases}
0,&(\forall~\Tc\subset\Yc^n)\ E(\xhat^n;\Tc)E(x^n;\Tc)
+E(\xhat^n;\Tc^c)E(x^n;\Tc^c)>0;\\
1,&\text{otherwise},
\end{cases}\\
&=&\begin{cases}
0,&\sum_{y^n\in\Yc^n}P_{Y^n|X^n}(y^n|\hat{x}^n)P_{Y^n|X^n}(y^n|x^n)>0;\\
1,&\text{otherwise},
\end{cases} \label{eqn:dmc_condition}
\end{IEEEeqnarray}
where $\Tc^c=\Yc^n\setminus\Tc$.  
We thus derive from Theorem~\ref{theorem2} that
\begin{IEEEeqnarray}{rCl}
C_0
&=&\sup_{n\geq 1}-\frac 1n\log \inf_{P_{X^n}}\Pr\left[\sum_{y^n\in\Yc^n}P_{Y^n|X^n}(y^n|\hat{X}^n)P_{Y^n|X^n}(y^n|X^n)>0\right]\\
&=&\sup_{n\geq 1}-\frac 1n\log \inf_{P_{X^n}}\Pr\left[\sum_{y^n\in\Yc^n}\left(\prod_{i=1}^nP_{Y|X}(y_i|\hat{X}_i)P_{Y|X}(y_i|X_i)\right)>0\right]\\
&=&\sup_{n\geq 1}-\frac 1n\log \inf_{P_{X^n}}\Pr\left[\prod_{i=1}^n
\sum_{y_i\in \Yc}P_{Y|X}(y_i|\hat{X}_i)P_{Y|X}(y_i|X_i)>0\right]\label{22}\\
&\geq&\sup_{n\geq 1}-\frac 1n\log \inf_{P_{X^n}\text{ i.i.d.}}\Pr\left[\prod_{i=1}^n
\sum_{y_i\in \Yc}P_{Y|X}(y_i|\hat{X}_i)P_{Y|X}(y_i|X_i)>0\right]\label{23}\\
&=&\sup_{n\geq 1}-\frac 1n\log \inf_{P_{X}}\left(\Pr\left[
\sum_{y\in \Yc}P_{Y|X}(y|\hat{X})P_{Y|X}(y|X)>0\right]\right)^n\\
&=&-\log \inf_{P_{X}}\Pr\left[
\sum_{y\in \Yc}P_{Y|X}(y|\hat{X})P_{Y|X}(y|X)>0\right].\label{25}
\end{IEEEeqnarray}
This concludes the proof of Corollary \ref{cor:dmc}.
\end{IEEEproof}


We remark that the inequality in \eqref{23} may be strict for DMCs. 
An example in which~\eqref{23} is a strict inequality  is given in~\cite[Fig.~2]{zero_error}, for which the five channel input letters form the vertices of a pentagon graph and an edge exists whenever $\mu(\hat x,x)=0$ as defined in \eqref{d-zero}. The zero-error capacities of the pentagon and some special generalizations
were later established by Lov\'asz in~\cite{lovasz}. However, for most other DMCs, the determination of the zero-error capacity remains a major  open problem in information theory and combinatorics. 


\section{Implications of the Distance Spectrum Formula for $M_n^\ast(d)$} \label{sec:implications}

In this section, further explorations based on the theoretical result in the previous section are conducted. 
Section~\ref{lower bound} shows that  the GV lower bound for discrete alphabets can be recovered from an information spectrum perspective. 
Section~\ref{subsec:UDbound} provides examples that leverage specifically 
uniform codeword  distributions. In particular, two of the examples considers   continuous code alphabets. 

\subsection{Relation to the Gilbert-Varshamov bound}
\label{lower bound}

An immediate consequence of Theorem~\ref{thm-m-size} is 
that a family of lower bounds to $M_n^\ast(d)$ can be  obtained by evaluating   $L_{X^n}(d)\triangleq 1/\Pr[\min\{\mu(\Xhat^n,X^n),\mu(X^n,\Xhat^n)\}<d]$ for different distributions $P_{X^n}$. This implies even if we do not use an optimal distribution $P_{X^n}$, we may still be able to obtain good lower bounds to the optimal code size.

From this perspective, the Gilbert-Varshamov (GV) lower bound~\cite{Moon05}
can be recovered with a uniform distribution over all possible codewords. 
Specifically, 
consider a finite code alphabet $\Xc$ with $|\Xc|=Q$ and the Hamming distance measure $\mu(\cdot,\cdot)$. 
Let the components of $X^n=(X_1\ X_2\ \ldots\ X_n)$ be i.i.d.\ and uniform over $\Xc$. This choice yields
\begin{IEEEeqnarray}{rCl}
\Pr\big[\mu(\Xhat^n, X^n)<d\big]=\Pr\bigg[\sum_{i=1}^n\mu(\hat X_i,X_i)<d\bigg]=\sum_{i=0}^{d-1}\binom ni \left(1- \frac{1}{ Q} \right)^i \left(\frac{1}{Q} \right)^{n-i}.\label{eq:betasum}
\end{IEEEeqnarray}
Hence, 
\begin{IEEEeqnarray}{rCl}
M_n^*(d)\geq\frac{1}{\sum_{i=0}^{d-1}\binom ni(1-\frac 1Q)^i(\frac 1Q)^{n-i}}=\frac{Q^n}{\sum_{i=0}^{d-1}\binom{n}i(Q-1)^{i}}\triangleq G_n(d),
\label{eq:DS-GV}
\end{IEEEeqnarray}
which is exactly the Gilbert-Varshamov (GV) lower bound  \cite{Moon05}.
The same observation has been stated by Kolesnik and Krachkovsky in~\cite[pp.~1446]{KK94}.

\subsection{Uniform Distribution (UD) Lower Bounds}
\label{subsec:UDbound}

The converse proof of Theorem~\ref{thm-m-size} shows that $M_n^\ast(d)$ can actually be achieved using a distribution which is {\em uniform} over an appropriate subset of $\Xc^n$ (that is, over an optimal code).
The discussion in the previous subsection also suggests that considering uniform $X^n$ 
over (a proper subset of) $\Xc^n$ may result in   good lower bounds. 
Thus, 
\begin{equation}
L_{X^n}(d)\triangleq\frac 1{\Pr\big[\min\{\mu(\Xhat^n,X^n),\mu(X^n,\Xhat^n)\}<d\big]}\end{equation}
based on uniform $P_{X^n}$ may form an important family of lower bounds
to $M_n^\ast(d)$.
For convenience, we will refer to this family of bounds as uniform distribution (UD) lower bounds.

\begin{example}\label{example1}
Here we derive an UD lower bound to $M_2^\ast(d)$
for Euclidean distance $\mu(\cdot,\cdot)$ and a bounded code alphabet $\Xc=[0,1)$.
Taking $P_{X^2}$ to be the uniform distribution over $\Xc^2$ 
and letting $Z_i\triangleq(\Xhat_i-X_i)^2$ yields that for $d>0$,
\begin{IEEEeqnarray}{rCl}
\Pr[\mu(\Xhat^2, X^2)<d]=\Pr\left[Z_1+Z_2<d^2\right]&=&\int_0^1\int_0^{d^2-z_1}f_Z(z_1)\, f_Z(z_2)\mathrm{d} z_2 \,\mathrm{d} z_1,
\end{IEEEeqnarray}
where
$f_Z(z)=\big(\frac 1{\sqrt{z}}-1\big){\bf 1}\{0\leq z<1\}$. We can check that
\begin{IEEEeqnarray}{rCl}
M_2^\ast(d)\geq \lceil L_{X^2}(d)\rceil 
=
\begin{cases}
3,&d=\frac 12;\\
2,&d=1.
\end{cases}\label{eq:ud-lower}
\end{IEEEeqnarray}
Via a procedure suggested by the proof of Theorem~\ref{thm-m-size}, we can actually obtain 
\begin{IEEEeqnarray}{rCl}
M_2^\ast(d)\geq\begin{cases}
8,&d=\frac 12;\\
2,&d=1. 
\end{cases} \label{eqn:num}
\end{IEEEeqnarray}
This indicates that 
there is room for improving 
the UD lower bound in \eqref{eq:ud-lower} in this situation
and the codeword selection procedure in the proof of Theorem~\ref{thm-m-size} could be further explored for finding a better lower bound.
\end{example}

\begin{example} 
We continue from the previous example but turn to a modified rectilinear distance measure $\mu(\xhat^n,x^n)\triangleq\sum_{i=1}^n
\min\{1-|\xhat_i-x_i|,|\xhat_i-x_i|\}$.
We again derive UD lower bounds to $M_2^\ast(d)$
for $\Xc=[0,1)$. Taking $P_{X^2}$ to be the uniform distribution over $\Xc^2$ 
and noting that each of $\mu(\Xhat_1,X_1)$ 
and $\mu(\Xhat_2,X_2)$ is uniform over $[0,1/2)$ and they are independent of each other, 
we have
\begin{IEEEeqnarray}{rCl}
\Pr[\mu(\Xhat^2, X^2)<d]
&=&\Pr\big[\mu(\Xhat_1,X_1)+\mu(\Xhat_2,X_2)<d\big]=\begin{cases}
2d^2,&0\leq d<\frac 12;\\
1-2(1-d)^2,&\frac 12\leq d<1,
\end{cases}
\end{IEEEeqnarray}
which implies
\begin{IEEEeqnarray}{rCl}
M_2^\ast(d)\geq L_{X^2}(d)=\begin{cases}
\frac 1{2d^2},&0\leq d<\frac 12;\\
\frac 1{1-2(1-d)^2},&\frac 12\leq d< 1.
\end{cases}
\label{eq:ex2-1}
\end{IEEEeqnarray}

To improve the above lower bound for $0\leq d\leq\frac 12$,
we adopt another distribution $P_{X^{2\circ}}$.
Let $X_1^\circ$ and $X_2^\circ$ be uniformly distributed over $\{0,d,2d,\ldots,(\ell-1)d\}$ and be independent of each other,
where $\ell\triangleq\lfloor 1/d\rfloor$. Then, $\Pr[\mu(\Xhat_1^\circ,X_1)
+\mu(\Xhat_2^\circ,X_2)<d]=\Pr[\Xhat_1^\circ=X_1]\Pr[\Xhat_2^\circ=X_2]=\frac 1{\ell^2}$,
and 
$M_2^\ast(d)\geq L_{X^{2\circ}}(d)=\big(\lfloor 1/d\rfloor\big)^2$.
Thus, when $1/d$ is an even integer, the lower bound in \eqref{eq:ex2-1} is doubled by taking a discrete distribution.

We can further improve $L_{X^{2\circ}}(d)$ for certain values of $d$ by introducing dependency among components. For $d=\frac 1k$ with $k\geq 2$, let $X_1^\ast$ be uniform over $\{0,\frac d2,d,\ldots,(2k-1)\frac d2\}$ and let $X_2^\ast$ be uniform over $\{X_1^\ast$, $(X_1^{\ast}+d)\,\text{mod}\,1$, $(X_1^{\ast}+2d)\,\text{mod}\,1$, $\ldots$,
$(X_1^{\ast}+(k-1)d)\,\text{mod}\,1\}$.
Thus, $X^{2\ast}$ is uniform over $2k^2$ points, each of which is at least at distance $1/k$ from all others. This implies 
$\Pr[\mu(\Xhat^{2\ast}, X^{2\ast})<\frac 1k]=\frac 1{2k^2}$,
and  
$M_2^\ast(\frac 1k)\geq L_{X^{2\ast}}(\frac 1k)=2k^2$, which improves $L_{X^{2\circ}}(1/k)=k^2$ for all integer $k\geq 2$.
By a similar technique to the Hamming bound, we can obtain
\begin{equation}
M^\ast_n\bigg(\frac 1k\bigg)\leq \frac{|\Xc|^2}{\min_{x^2 \in \Xc^2} |\{\xhat^2\in\Xc^2:
\mu(\xhat^2,x^2)<\frac 1{2k}\}|}=\frac 1{\frac 1{2k^2}}=2k^2,
\end{equation}
where $|\cdot|$ here denotes the area in a 2-dimensional real plane.
Therefore, $L_{X^{2\ast}}(1/k)=M_2^\ast(1/k)$ for $k\geq 2$ is indeed tight.
\color{black}

For $\frac 12\leq d< 1$, we introduce $P_{X^{2\diamond}}$ with strongly dependent components as $X_2^\diamond=X_1^\diamond$ with probability one. 
Then, noting that  $\mu(\Xhat_2^\diamond,X_2^\diamond)$ is equal to $\mu(\Xhat_1^\diamond,X_1^\diamond)$ with probability one,
we have $\Pr[\mu(\Xhat^{2\diamond}, X^{2\diamond})<d]
=\Pr[\mu(\Xhat_1^\diamond,X_1^\diamond)+\mu(\Xhat_2^\diamond,X_2^\diamond)<d]=\Pr[\mu(\Xhat_1^\diamond,X_1^\diamond)<\frac d2]$.
With $X_1^\diamond$ uniform over $\{0,\frac d2,\ldots,(j-1)\frac d2\}$, where $j\triangleq\lfloor 2/d\rfloor$, we obtain
$\Pr[\mu(\Xhat_1^\diamond,X_1^\diamond)<\frac d2]=\Pr[\Xhat_1^\diamond=X_1^\diamond]=1/j$, which implies
$M_2^\ast(d)\geq L_{X^{2\diamond}}(d)=\lfloor 2/d\rfloor$.
Accordingly, when $\frac 12<d\leq \frac 23$, $L_{X^{2\diamond}}(d)=3>\lceil L_{X^{2}}(d)\rceil =2$, and hence improvement can be obtained
by introducing dependence among components.
\end{example}

\begin{example} 
In this example, we demonstrate a case that $M_n^\ast(d)$ can be exactly determined and hence its relation with the UD lower bound $L_{X^n}(d)$ can be quantitatively examined.

Let the distance measure be given by
\begin{IEEEeqnarray}{rCl}
\mu(\xhat^n, x^n)=|\kappa_n(\xhat^n)-\kappa_n(x^n)|,
\end{IEEEeqnarray}
where $\xhat^n$ and $x^n$ are in $\{0,1\}^n$, and $\kappa_n(x^n)\triangleq x_n2^{n-1}+x_{n-1}2^{n-2}+...+x_22^1+x_1$
is the binary representation of $x^n=(x_1\ x_2\ \ldots\ x_n)$.
In other words, $\mu(\xhat^n,x^n)$
is the absolute difference between two decimal numbers
$\kappa_n(\xhat^n)$ and $\kappa_n(x^n)$, and is 
a \emph{separable distance measure} \cite[Def.~1]{CLH00}.

Since $\kappa_n(x^n)$ is an integer in $\{0,1,2,\ldots,2^n-1\}$,
it can be easily seen that for $d>0$,
\begin{IEEEeqnarray}{rCl}
M_n^\ast(d)=\left \lceil {\frac{2^n}{\lceil d \rceil}} \right \rceil,
\end{IEEEeqnarray}
where $\lceil\cdot\rceil$ is the ceiling function.
Notably, one of the uniform $X^n$'s that results in $L_{X^n}(d)=M_n^\ast(d)$ has support $\{0,\lceil d \rceil,2\lceil d \rceil,\ldots,(M_n^\ast(d)-1)\lceil d \rceil\}$, and there are exactly $\lceil d \rceil$
optimizers that can achieve $M_n^\ast(d)$.

We then recall that \eqref{eq:DS-GV} has illustrated that $G_n(d)$ can be regarded as a special case of the UD lower bound with uniform $X^n$ over the entire $\Xc^n$. As such, we derive
\begin{IEEEeqnarray}{rCl}
G_n(d)&=&\frac 1{\Pr\{|\kappa_n(\Xhat^n)-\kappa_n(X^n)|<d\}}=\frac 1{\Pr\{|\kappa_n(\Xhat^n)-\kappa_n(X^n)|<\lceil d \rceil \}}\\[2mm]
&=&\begin{cases}
\displaystyle\frac{2^{2n}}{(3\lceil d \rceil-1)\lceil d \rceil+(2\lceil d \rceil-1)(2^n-2\lceil d \rceil)}, &0<\lceil d \rceil\leq 2^{n-1}\\[3mm]
\displaystyle\frac{2^{2n}}{2^{2n}+(\lceil d\rceil-2^n)(2^n-\lceil d\rceil +1)},& 2^{n-1}<\lceil d\rceil\leq 2^n-1 ;\\[3mm]
1,&\lceil d \rceil>2^n-1,
\end{cases}
\end{IEEEeqnarray}
showing that $G_n(d)$ is strictly less than $M_n^\ast(d)$ 
except when $\lceil d\rceil=1$ and $\lceil d\rceil\geq 2^n$.
This result confirms that the finite length GV lower bound is not tight in general.

We close this example by noting that an upper bound $U_n(d)$ for $M_n^\ast(d)$ can also be provided based on Theorem~\ref{thm-m-size}. 
If there exists $U_n(d)$ such that $U_n(d)\geq 1/\Pr{[\min\{\mu(\Xhat^n,X^n),\mu(X^n,\Xhat^n)\}<d}]$ for all $P_{X^n}$'s, then
\begin{IEEEeqnarray}{rCl}
U_n(d)\geq\frac 1{\inf_{P_{X^n}}\Pr{[\min\{\mu(\Xhat^n,X^n),\mu(X^n,\Xhat^n)\}<d}]}=M_n^\ast(d).
\end{IEEEeqnarray}
Now setting $j=j(n,d)\triangleq 2^n/\lceil d \rceil $, we derive
\begin{IEEEeqnarray}{rCl}
\lefteqn{
\Pr\left\{\bigg|\frac{\kappa_n(\Xhat^n)}{2^n}-
\frac{\kappa_n(X^n)}{2^n}\bigg|<\frac{\lceil d \rceil}{2^n}\right\}}\\
&=&\Pr\left\{\bigg|\frac{\kappa_n(\Xhat^n)}{2^n}-
\frac{\kappa_n(X^n)}{2^n}\bigg|<\frac 1j\right\}\\
&\geq&\sum_{i=0}^{\lceil j\rceil-1}\Pr\left\{
\frac i{\lceil j\rceil}\leq\frac{\kappa_n(\Xhat^n)}{2^n}<\frac{i+1}{\lceil j\rceil}
\text{ and }\frac i{\lceil j\rceil}\leq\frac{\kappa_n(X^n)}{2^n}<\frac{i+1}{\lceil j\rceil}\right\}\\
&=&\sum_{i=0}^{\lceil j\rceil-1}\Pr\left\{
\frac i{\lceil j\rceil}\leq\frac{\kappa_n(\Xhat^n)}{2^n}<\frac{i+1}{\lceil j\rceil}\right\}\Pr\left\{\frac i{\lceil j\rceil}\leq\frac{\kappa_n(X^n)}{2^n}<\frac{i+1}{\lceil j\rceil}\right\}\label{eq:i.i.d.-1}\\
&=&\sum_{i=0}^{\lceil j\rceil-1}\left(\Pr\left\{\frac i{\lceil j\rceil}\leq\frac{\kappa_n(X^n)}{2^n}<\frac{i+1}{\lceil j\rceil}\right\}\right)^2\label{eq:i.i.d.-2}\\
&\geq&\frac{1}{\lceil j\rceil},\label{eq:CS}
\end{IEEEeqnarray}
where \eqref{eq:i.i.d.-1} and \eqref{eq:i.i.d.-2} hold because 
$\Xhat^n$ and $X^n$ are i.i.d.,
and \eqref{eq:CS} again follows from the Cauchy-Schwarz inequality.
This immediately gives 
\begin{IEEEeqnarray}{rCl}
U_n(d)=\lceil j\rceil=\left\lceil \frac{2^n}{\lceil d \rceil} \right\rceil,
\end{IEEEeqnarray}
which is precisely $M_n^\ast(d)$.
\hfill$\Box$
\end{example}

\section{Extensions to the Asymptotic Regime}\label{sec5:asym}

We now extend the result in Theorems~\ref{thm-m-size} to the asymptotic regime in which the length $n$ of the code goes to infinity. In what follows, $\log$ denotes the natural logarithm. A distance spectrum formula for the largest code rate $R=\log(M)/n$ subject to a normalized minimum distance $\delta=d/n$ can be obtained on the basis of Theorem~\ref{thm-m-size} in a straightforward manner:
\begin{IEEEeqnarray}{rCl}
R_n^\ast(\delta)&\triangleq&\frac 1n\log M_n^\ast(n\delta)=\sup_{P_{X^n}}\left(-\frac 1n\log\Pr\left[\frac 1n\mu(\Xhat^n,X^n)<\delta\right]\right).\label{eq:rate}
\end{IEEEeqnarray}
The formula of $R^\ast_n(\delta)$ in \eqref{eq:rate} provides a quantitative characterization of the largest code rate attainable for an $(n,M,n\delta)$-code,
based on which a first-order expression for the largest asymptotic code rate attainable for a sequence of  $(n,M,n\delta)$-codes can be obtained when
the normalized distance measure is uniformly bounded.


\begin{theorem}\label{asymptotic} 
{\it (Largest Asymptotic Code Rate)} 
Fix an arbitrary code alphabet $\Xc$ 
and a (sequence of) general distance measures $\mu(\cdot,\cdot)$ that satisfy the condition mentioned in Theorem \ref{thm-m-size} and also satisfy
\begin{equation}
\sup_{n\geq 1}\max_{\xhat^n,x^n\in\Xc^n}\frac 1n\mu(\xhat^n,x^n)<\infty.
\label{eq:uniform-bounded}
\end{equation}
Then,
\begin{IEEEeqnarray}{rCl}
\limsup_{n\rightarrow\infty} R_n^\ast(\delta) &=&\limsup_{n\rightarrow\infty}\sup_{P_{X^n}}J_{X^n}(\delta)
\quad\text{and}\\
\liminf_{n\rightarrow\infty} R_n^\ast(\delta)&=&\liminf_{n\rightarrow\infty}\sup_{P_{X^n}}J_{X^n}(\delta),
\end{IEEEeqnarray}
where 
\begin{IEEEeqnarray}{rCl}J_{X^n}(\delta)\triangleq \inf_{a\leq \delta}\sup_{\theta\in\Re}\left\{a\theta-\frac 1n\log\E\left[e^{\theta\mu(\Xhat^n,X^n)}\right]\right\}.\label{J}
\end{IEEEeqnarray} 
\end{theorem}
\begin{IEEEproof} 
The proof can be found in Appendix~\ref{AA}. In particular, an upper bound on the 
 second-order term  of $R_n^\ast(\delta)$ 
is also provided (cf.\ Lemma~\ref{lemma2}). 
\end{IEEEproof}


The above theorem indicates that 
$R_n^\ast(\delta)$ and $\sup_{P_{X^n}}J_{X^n}(\delta)$ are asymptotically close.
In fact, the proof in Appendix A shows that $R_n^\ast(\delta)\geq\sup_{P_{X^n}}J_{X^n}(\delta)$ for every $n$. However, the proof of the upper bound is significantly more involved and requires delicate twisting of probability distributions~\cite{chen00}. 

A lower bound to $R_n^\ast(\delta)$ can be 
obtained when $P_{X^n}$ is i.i.d.~and $\mu(\cdot,\cdot)$ is additive, e.g.,  the Hamming distance.


\begin{corollary}\label{corollary} 
Assume that $\Xc=\{\alpha_1,\alpha_2,...\alpha_Q\}^n$ and $\mu(\cdot,\cdot)$ is the Hamming distance measure. For any $\delta >0$, one has 
\begin{IEEEeqnarray}{rCl}
R_n^\ast(\delta)&\geq&\begin{cases}
D\left(\delta\left\|\frac{Q-1}{Q}\right.\right),&0<\delta<\frac{Q-1}{Q};\\
0,&\delta\geq \frac{Q-1}{Q},
\end{cases}\label{eq:AVGbound}
\end{IEEEeqnarray}
where 
\begin{equation}
D(a\|b)\triangleq a\log\left(\frac{a}{b}\right)+(1-a)\log\left(\frac{1-a}{1-b}\right)
\end{equation} is the binary Kullback-Leibler divergence \cite{CT06}. 
\end{corollary}
\begin{IEEEproof} The proof is deferred to Appendix~\ref{AB}.
\end{IEEEproof}

Using a large deviations technique, we can slightly improve \eqref{eq:AVGbound} by the addition of a logarithmic term as 
for $Q=2$ and $0<\delta < 1/2$, 
\begin{equation}
R_n^\ast(\delta)\ge D\left( \delta \, \middle\|\,\frac{1}{2}\right) + \frac{\log n }{2n} +  \Theta\bigg(\frac{1}{n}\bigg)\quad\text{as }n\to\infty.\label{eqn:Rn_improve}
\end{equation}
Although Jiang and Vardy~\cite[Thm.~1]{JiangVardy04}  have shown, by using a graph-theoretic framework, that  the achievable second-order term in \eqref{eqn:Rn_improve} is at least $(\log n)/n$, which is slightly stronger than the term $(\log n)/(2n)$, Eq.~\eqref{eqn:Rn_improve} provides some additional insight into the suboptimality of choosing $\Xhat^n$ and $X^n$ with i.i.d.\ components (since our evaluation of the relevant distance spectrum is asymptotically tight when employing distributions to yield dependent elements of $\Xhat^n$ and $X^n$). 

\section{Conclusion and Future Work}
\label{sec:conclusion}

In this paper, we developed an exact formula for the maximal size of distance-$d$ codes for arbitrary alphabets and general distance measures. The implications of the established formula were discussed. The extension to the asymptotic regime was also explored. Some natural directions for future work includes: 
\begin{itemize}
\item Understanding the structure of optimal or even ``good'' distributions $P_{X^n}$ to give lower bounds on the optimal code size. For example, based on our numerical experiments, we know that the optimal distribution may not be unique. Studying the binary Hamming distance for small block lengths suggests that there may be an optimizer whose marginals are uniform on each coordinate. 
\item Seek   $i)$ a similar formula of the minimum code size subject to a covering radius constraint (cf.\ \cite{chen01}) and $ii)$ a formula of maximal code size under a minimum multi-wise distance constraint (cf.\ \cite{LinMoser18}). The latter would constitute   a generalization of Tur\'an's Theorem.
\end{itemize}

\appendices
\section{Proof of Theorem~\ref{asymptotic}}
\label{AA}

The theorem can be verified via the following two lemmas.
The first lemma
shows that for arbitrary distance measures, $R_n^\ast(\delta)$ is lower-bounded by $\sup_{P_{X^n}}J_{X^n}(\delta)$. The second lemma proves that 
$R_n^\ast(\delta)$ is upper bounded by $\sup_{X^n}J_{X^n}(\delta) + \Theta( \frac{1}{\sqrt{n}})$ when the normalized distance measure is uniformly bounded.
Then, the two lemmas imply Theorem~\ref{asymptotic}.

\begin{lemma}\label{lemma1}
Fix an arbitrary code alphabet and an arbitrary distance measure that 
satisfies \eqref{dist-cond}. Then,
\begin{IEEEeqnarray}{rCl}
R_n^\ast(\delta)\geq \sup_{P_{X^n}}J_{X^n}(\delta).
\label{eq:T4}
\end{IEEEeqnarray}
\end{lemma}
\begin{IEEEproof}
It suffices to prove that for every $P_{X^n}$,
\begin{equation}
-\frac 1n\log\Pr\left[\frac 1n\mu(\Xhat^n,X^n)<\delta\right]\geq J_{X^n}(\delta).
\label{79}
\end{equation}
Denote 
\begin{equation}\label{eq:ldrf}
I_{X^n}(a)\triangleq\sup_{\theta\in\Re}\left\{a\theta-\varphi_{X^n}(\theta)\right\}
\end{equation}
as the large deviation rate function with respect to the normalized random distance $\frac 1n\mu(\Xhat^n,X^n)$, 
where 
$\varphi_{X^n}(\theta)\triangleq\frac 1n\log\E[e^{\theta\mu(\Xhat^n,X^n)}]$.
An elementary property\footnote{
The large deviation rate function 
$I_{X^n}(a)$ is convex, and admits its global minimum $\min_{a\in\Re}I_{X^n}(a)=0$ at $a=(1/n)\E[\mu(\Xhat^n,X^n)]$ \cite{Bucklew90}.} of the large deviation rate function 
gives that
\begin{IEEEeqnarray}{rCl}
J_{X^n}(\delta)=\begin{cases}
I_{X^n}(\delta),&\delta<\displaystyle\mbox{$\frac 1n$}\E[\mu(\Xhat^n,X^n)];\\
0,&\text{otherwise}.
\end{cases}\label{eq:large}
\end{IEEEeqnarray}
Thus, it suffices to prove \eqref{79} under the condition that 
$\delta<\frac 1n\E[\mu(\Xhat^n,X^n)]$ (since $0$ is a trivial lower bound).
Let $Y\triangleq n\delta-\mu(\Xhat^n,X^n)$ and note from the previous condition 
that $\E[Y]<0$. We then derive from Markov's inequality that for $\theta>0$,
\begin{IEEEeqnarray}{rCl}
\Pr\left[\frac 1n\mu(\Xhat^n,X^n)<\delta\right]&=&
\Pr[Y>0]=\Pr[e^{\theta Y}>1]\leq\E[e^{\theta Y}]\triangleq M_Y(\theta). \label{eqn:markov}
\end{IEEEeqnarray}
Applying the fact that $\left.\frac{\partial}{\partial \theta}M_Y(\theta)\right|_{\theta=0}= \mathbb{E}[Y]<0$ and the convexity of $M_Y(\theta)$ over $\theta\in\Re$, we obtain
\begin{IEEEeqnarray}{rCl}
\Pr\left[Y>0\right]
&\leq&\inf_{\theta>0}M_Y(\theta)=\inf_{\theta\in\Re}M_Y(\theta)
=\inf_{\theta\in\Re}M_Y(-\theta)\\
&=&\exp\left\{-n\sup_{\theta\in\Re}\left(\delta\theta-\frac 1n\log\E\left[e^{\theta\mu(\Xhat^n,X^n)}\right]\right)\right\}\\
&=&\exp\left\{-n\cdot I_{X^n}(\delta)\right\},
\end{IEEEeqnarray}
which complete the proof of \eqref{79}.
\end{IEEEproof}


\begin{lemma}\label{lemma2} 
Fix an arbitrary code alphabet and an arbitrary distance measure that 
satisfies both \eqref{dist-cond} and 
\eqref{eq:uniform-bounded}. Then, given that
$P_{X^n}$ is the optimizer of $\sup_{X^n}J_{X^n}(\delta)$, we have
\begin{IEEEeqnarray}{rCl}
R_n^\ast(\delta)&\leq& I_{X^n}(\delta)+\frac 4{\sqrt{n}}\left[\frac 1n
\frac{\varphi_{X^n}^{(4)}(-\theta^\ast)}{(\varphi_{X^n}''(-\theta^\ast))^2}+3\right]\sqrt{\varphi_{X^n}''(-\theta^\ast)}\nonumber\\
&&+\frac 1n\log \left(\frac 4n
\frac{\varphi_{X^n}^{(4)}(-\theta^\ast)}{(\varphi_{X^n}''(-\theta^\ast))^2}+12\right)\label{eq:T5}
\end{IEEEeqnarray}
for those $\delta$ satisfying $\sup_{P_{X^n}}J_{X^n}(\delta)>0$.
\end{lemma}
\begin{IEEEproof} 
Following the notations used in the proof of Lemma~\ref{lemma1}, we define the twisted distribution of $Y$ as
\begin{equation}
\mathrm{d}P_{Y^{(\theta)}}(y)\triangleq\frac{e^{\theta y} \, \mathrm{d} P_Y(y)}{M_Y(\theta)}.
\end{equation}
Then, 
\begin{IEEEeqnarray}{rCl}
\Pr[Y>0]&=&\int_0^\infty \, \mathrm{d}P_Y(y)\\
&=&\int_0^\infty M_Y(\theta^\ast)e^{-\theta^\ast y}\, \mathrm{d}P_{Y^{(\theta^\ast)}}\\
&=&M_Y(\theta^\ast)\int_0^\infty 
e^{-\theta^\ast y}\, \mathrm{d}P_{Y^{(\theta^\ast)}}(y),\label{converse1}
\end{IEEEeqnarray}
where $\theta^\ast$ is the minimizer of $\inf_{\theta\in\Re}M_Y(\theta)$.
Note that since $P_{X^n}$ is the optimizer for $\sup_{P_{X^n}}J_{X^n}(\delta)$ 
and only those $\delta$ satisfying $\sup_{P_{X^n}}J_{X^n}(\delta)>0$ is considered,
we can infer from \eqref{eq:large} that $\E[Y]<0$ and hence $0<\theta^\ast<\infty$.
By noting that $\Pr[Y^{(\theta)}> 0]$ is positive,\footnote{\label{footnote6}
If $\Pr[Y^{(\theta^\ast)}>0]=0$, then $-Y^{(\theta^\ast)}$ is a non-negative random variable. We can therefore apply Markov's inequality to obtain  $\Pr[-Y^{(\theta^\ast)}\geq\epsilon]\leq\E[-Y^{(\theta^\ast)}]/\epsilon=0$ 
since the minimizer of $\inf_{\theta\in\Re} M_Y(\theta)$ must validate  $E[Y^{(\theta^\ast)}]=0$ \cite{Bucklew90}.
As a result, $\Pr[-\epsilon<Y^{(\theta^\ast)}\leq 0]=1$ for arbitrary $\epsilon>0$,
which implies $\Pr[Y^{(\theta^\ast)}=0]=1$. 
As $Y^{(\theta^\ast)}$ and $Y$ have the same support, we conclude $\Pr[Y=0]=1$,
thereby resulting a contradiction to $\E[Y]<0$.} 
we let $W$ be a nonnegative random variable with distribution
\begin{IEEEeqnarray}{rCl}
  \mathrm{d}P_W(y)\triangleq\frac{  \mathrm{d}P_{Y^{(\theta^\ast)}}(y)}{\Pr[Y^{(\theta)}> 0]}.
\end{IEEEeqnarray}
Then, \eqref{converse1} can be rewritten as
\begin{IEEEeqnarray}{rCl}
\Pr[Y> 0]&=&M_Y(\theta^\ast)\int_0^\infty e^{-\theta^\ast y}\, \mathrm{d}P_{Y^{(\theta^\ast)}}(y)\\
&=&M_Y(\theta^\ast)\cdot\Pr[Y^{(\theta^\ast)}> 0]
\int_0^\infty e^{-\theta^\ast y}\, \mathrm{d}P_{W}(y)\\
&=&M_Y(\theta^\ast)\cdot\Pr[Y^{(\theta^\ast)}> 0]
\cdot\E\left[e^{-\theta^\ast W}\right].
\end{IEEEeqnarray}
Using the fact that $\E[Y^{(\theta^\ast)}]=0$ \cite[Thm.~9.2]{Bucklew90}, we obtain\footnote{
$\Pr[Y^{(\theta^\ast)}>0]>0$ and $\E[Y^{(\theta^\ast)}]=0$ jointly imply $\E\{[Y^{(\theta^\ast)}]^2\}>0$
and $\E\{[Y^{(\theta^\ast)}]^4\}>0$, which justifies \eqref{thm9-2}.
}
\begin{IEEEeqnarray}{rCl}
\Pr[Y^{(\theta^\ast)}> 0]&\geq&\frac{\E^2[(Y^{(\theta^\ast)})^2]}
{4\,\E[(Y^{(\theta^\ast)})^4]}\label{thm9-2}\\
&=&\frac{(M_Y''(\theta^\ast))^2}{
4M_Y(\theta^\ast)M_Y^{(4)}(\theta^\ast)}\\
&=&\frac{(C_Y''(\theta^\ast))^2}{4\left[C_Y^{(4)}(\theta^\ast)+3(C_Y''(\theta^\ast))^2\right]}\\
&=&\frac{1}{4\Big[\frac 1n
\frac{\varphi_{X^n}^{(4)}(-\theta^\ast)}{(\varphi_{X^n}''(-\theta^\ast))^2}+3\Big]}
\end{IEEEeqnarray}
where 
\begin{IEEEeqnarray}{rCl}
C_Y(\theta)  \triangleq  \log M_Y(\theta)&=& n\theta \delta+\log\E[e^{-\theta \mu(\hat X^n,X^n)}]=n\left(\theta \delta+\varphi_{X^n}(-\theta)\right),\\
C_Y''(\theta) &=&  n\cdot\varphi_{X^n}''(-\theta),
\end{IEEEeqnarray}
 and
 \begin{equation}
 C_Y^{(4)}(\theta)=n\cdot\varphi_{X^n}^{(4)}(-\theta).
 \end{equation}
Notably, for a bounded distance measure $\mu(\cdot,\cdot)$, $\varphi_{X^n}(\theta)$ is guaranteed to be fourth-order differentiable.
Using Jensen's inequality, we derive
\begin{IEEEeqnarray}{rCl}
\E\left[e^{-\theta^\ast W}\right]&\geq&e^{-\theta^\ast\cdot \E[W]},
\end{IEEEeqnarray}
and hence 
\begin{IEEEeqnarray}{rCl}
\E[W]&=&\int_0^\infty w\, \mathrm{d}P_W(w)\\
&=&\int_0^\infty y\frac{  \mathrm{d}P_{Y^{(\theta^\ast)}}(y)}{\Pr[Y^{(\theta^\ast)}> 0]}\\
&\leq&\frac 1{\Pr[Y^{(\theta^\ast)}> 0]}\int_{-\infty}^\infty |y|\, \mathrm{d}P_{Y^{(\theta^\ast)}}(y)\\
&=&\frac 1{\Pr[Y^{(\theta^\ast)}>0]}\E[|Y^{(\theta^\ast)}|]\\
&\leq&\frac 1{\Pr[Y^{(\theta^\ast)}> 0]}\sqrt{\E[(Y^{(\theta^\ast)})^2]}\\
&=&\frac 1{\Pr[Y^{(\theta^\ast)}> 0]}\sqrt{\frac{M_Y''(\theta^\ast)}{M_Y(\theta^\ast)}}\\
&=&\frac 1{\Pr[Y^{(\theta^\ast)}> 0]}\sqrt{C_Y''(\theta^\ast)}\\
&=&\frac 1{\Pr[Y^{(\theta^\ast)}> 0]}\sqrt{n\cdot\varphi_{X^n}''(-\theta^\ast)}.
\end{IEEEeqnarray}
We conclude from all the above derivations that
\begin{IEEEeqnarray}{rCl}
\Pr[Y>0]&\geq&e^{-n\cdot I_{X^n}(\delta)}
\frac{e^{- 4\Big(\frac 1n
\frac{\varphi_{X^n}^{(4)}(-\theta^\ast)}{(\varphi_{X^n}''(-\theta^\ast))^2}+3\Big)\sqrt{n\cdot\varphi_{X^n}''(-\theta^\ast)}}}{4\Big(\frac 1n
\frac{\varphi_{X^n}^{(4)}(-\theta^\ast)}{(\varphi_{X^n}''(-\theta^\ast))^2}+3\Big)},
\end{IEEEeqnarray}
which concludes the proof of \eqref{eq:T5}.
\end{IEEEproof}

We complete the proof of Theorem~\ref{asymptotic}
by remarking that
with probability one, 
$(1/n)\mu(\Xhat^n,X^n)$ is not only bounded, but uniformly upper bounded in the block length $n$, and so are its moments and cumulants. Since a twisted random variable generated from $(1/n)\mu(\Xhat^n,X^n)$
must have the same support as $(1/n)\mu(\Xhat^n,X^n)$, its twisted moments as well as twist cumulants are also uniformly bounded. Accordingly, 
$\varphi_{X^n}^{(4)}(-\theta^\ast)=O(1)$ and 
$\varphi_{X^n}''(-\theta^\ast)=O(1)$, based on which 
\eqref{eq:T5} implies 
$R_n^\ast(\delta)\leq\sup_{X^n}J_{X^n}(\delta) + \Theta( \frac{1}{\sqrt{n}})$.

\section{Proof of Corollary~\ref{corollary}}
\label{AB}
\begin{IEEEproof}
When the distance measure $\mu(\cdot,\cdot)$ is additive and the components of $X^n$ are i.i.d.\ with generic distribution $P_X$, $I_{X^n}(a)$ 
as defined in \eqref{eq:ldrf} exhibits a single-letter expression for all block lengths $n$ as:
\begin{IEEEeqnarray}{rCl}
I_{X^n}(a)=I_X(a)=\sup_{\theta\in\Re}\left\{a\theta-\varphi_{X}(\theta)\right\},
\end{IEEEeqnarray}
where 
$\varphi_{X}(\theta)\triangleq \log\E\big[e^{\theta\mu(\hat X,X)}\big].$
Thus, when  $X^n$ has i.i.d. components, \eqref{J} is simplified to
\begin{IEEEeqnarray}{rCl}J_{X^n}(\delta)\triangleq \inf_{a\leq \delta}I_{X^n}(a)=\inf_{a\leq \delta}I_{X}(a)\triangleq J_{X}(\delta).
\end{IEEEeqnarray} 
This and \eqref{eq:T4} lead to
\begin{IEEEeqnarray}{rCl}
R_n^\ast(\delta)&\geq&\sup_{P_{X^n}} J_{X^n}(a)\geq \sup_{P_{X^n}\text { i.i.d.}} J_{X^n}(a)=\sup_X J_X(a).\label{R_n}
\end{IEEEeqnarray}
As long as $\Xc=\{\alpha_1,\alpha_2,\ldots,\alpha_{Q}\}$ is finite and $\mu(\cdot,\cdot)$ is the Hamming distance measure, we have
\begin{IEEEeqnarray}{rCl}
\varphi_{X}(\theta)&=&\log
\left(\sum_{i=1}^{Q}\sum_{j=1}^{Q}P_{X}(\alpha_i)P_{X}(\alpha_j)e^{\theta\mu(\alpha_i,\alpha_j)}\right)\\
&=&\log\left(1-b_X+b_Xe^{\theta}\right),
\end{IEEEeqnarray}
and
\begin{IEEEeqnarray}{rCl}
I_X(a)&=&\sup_{\theta\in\Re}\left\{a\theta-\varphi_{X}(\theta)\right\}=D(a\|b_X),\label{I_X}
\end{IEEEeqnarray}
where $b_X\triangleq 1-\sum_{i=1}^{Q}P_{X}^2(\alpha_i)=\E[\mu(\Xhat,X)]$. Again, according to the property of the large deviation rate function \cite{Bucklew90}, we have 
\begin{IEEEeqnarray}{rCl}
J_X(\delta)=\begin{cases}
I_X(\delta),&0<\delta<b_X;\\
0,&\text{otherwise}.\label{J_X}
\end{cases}
\end{IEEEeqnarray}
According to the Cauchy-Schwarz inequality, 
$0\leq b_X\leq 1-1/Q=(Q-1)/Q$.
Combining \eqref{R_n}, \eqref{I_X} and \eqref{J_X}, we conclude that
for $\delta>0$,
\begin{IEEEeqnarray}{rCl}
R_n^\ast(\delta)&\geq&\sup_X J_X(\delta)=\sup_{0\leq b_X\leq (Q-1)/Q}J_X(\delta)=\begin{cases}
 D\left(\delta\left\|\frac{Q-1}{Q}\right.\right),&0<\delta<\frac{Q-1}{Q};\\
0,&\delta\geq \frac{Q-1}{Q}.
\end{cases}
\end{IEEEeqnarray}
\end{IEEEproof}

\subsection*{Acknowledgements}

The work of Ling-Hua Chang and Po-Ning Chen is supported by 
the Ministry of Science and Technology (MoST), Taiwan, under grant 105-2221-E-009-009-MY3. 
The work of Carol Wang and Vincent Tan is supported by a Singapore Ministry of Education (MoE) Tier 2 grant (R-263-000-B61-112). The work of Yunghsiang S. Han is support by the National Natural Science Foundation of China (Grant No. 61671007).

A special acknowledgement
is given to Dr.~Mladen Kova\v{c}evi\'{c}, 
who brought 
\cite{KK94} to the authors' attention.
The authors sincerely thank the anonymous reviewers for their constructive feedback to improve the quality of the paper.
\bibliographystyle{IEEEtran}

\bibliography{biblio_dis_pn}

\end{document}